\documentstyle[aps, preprint]{revtex}
\begin{document}
\tightenlines
\preprint{}
\title{Exploring CP violation with $B_s^0 \to \bar{D^0} \phi $ decays}
\author{A. K. Giri$^1$, R. Mohanta$^2$ and M. P. Khanna$^1$}
\address{ $^1$ Physics Department, Panjab University, Chandigarh-160 014,
India}
\address{ $^2$ School of Physics, University of Hyderabad,
Hyderabad-500 046, India }
\maketitle
\begin{abstract}
We note that it is possible to determine the weak phase $\gamma $
from the time dependent measurements of the decays
$B_s^0(t) (\bar B_s^0 (t)) \to \bar D^0 \phi $ without any
hadronic uncertainties. These decays are described by the color
suppressed tree diagrams and hence are free from the
penguin pollutions. We further demonstrate that $\gamma$ can also
be extracted with no hadronic uncertainties from an angular analysis of
corresponding vector vector modes,
$B_s^0(t) (\bar B_s^0 (t)) \to \bar D^{* 0} \phi $. Although the branching
ratios for these decay modes are quite small $ {\cal O} (10^{-5}-10^{-6})$,
the strategies presented here
appear to  be particularly interesting for the ``second generation" experiments
at hadronic $B$ factories.
\end{abstract}


\section{Introduction}

Despite many attempts, CP violation still remains one of the
most outstanding problems in particle physics \cite{kog89,bigi00,lav99}.
The standard model (SM) with three generations provides a simple
description of this phenomenon through the complex
Cabibbo-Kabayashi-Maskawa (CKM) matrix \cite{ref1}. Decays of $B$ mesons
provide rich ground for investigating $CP$ violation
\cite{buras98,quin98}. They allow  stringent tests both for the 
SM and for studies
of new sources of this effect. Within the SM, the $CP$ violation is
often characterized by the so-called unitarity triangle \cite{chau88}.
Detection of CP violation and the accurate determination of the
unitarity triangle are the major goals of experimental $B$
physics \cite{stone94}. Decisive information about the origin
of CP violation in the flavor sector can  be obtained if the three angles
$\alpha (\equiv \phi_2)$, $\beta(\equiv \phi_1)$
and $\gamma(\equiv \phi_3)$ can be independently measured
\cite{bigi91}. Within the Standard Model the sum of these
three angles is equal to
$180^\circ $. Thus one tests the Standard Model by testing whether
independent  determination of the three angles give consistent results.
Over the past decade or so many methods have been proposed for
obtaining the three interior angles of the unitarity triangle. In the
near future  these CP phases will be measured in a variety of experiments
at $B$ factories, HERA-B and hadron colliders.

The CP angles are typically
extracted from CP violating rate asymmetries in $B$ decays
\cite{quin}.
The phase $ \beta \equiv {\rm Arg} (-V_{cb}^* V_{cd}/V_{tb}^* V_{td})$
(=Arg($V_{td}^*$), in the standard phase convention) is measured by the
time dependent CP asymmetry in $B_d^0(t) \to J/\psi K_s $. Theoretically,
this measurement provides a very clean determination of $\sin 2 \beta$,
since the single phase approximation holds in this case
\cite{ref33,ref34}. Recently, large CP violation in $B_d^0 \to
\psi K_s $ has been observed by Babar and Belle Collaborations
and clean measurement of $\beta $ has been made
\cite{babar01,belle01}. This is the first step
towards a serious test of standard model of CP violation.

The angle $\alpha \equiv {\rm Arg}(-V_{tb}^*
V_{td}/V_{ub}^* V_{ud})$ can be measured using the CP
asymmetries in the decays $B_d^0 \to \pi^+ \pi^- $, however there are
theoretical hadronic uncertainties due to the existence of penguin diagrams
\cite{ref33,ref34}. A theoretical cleaner way of resolving the penguin
correction will require the combination of  the asymmetry in $ B_d^0 \to \pi^+ \pi^- $
with other measurements. A very early suggestion \cite{ref35} was that one also has to 
measure the isospin related processes $B^+ \to \pi^+ \pi^0 $ and
$B^0/\bar{B^0} \to \pi^0 \pi^0 $ and can thereby extract the angle $\alpha $
with reasonable accuracy.

The most difficult to measure is the angle $\gamma \equiv
{\rm Arg} (-V_{ub}^* V_{ud}/V_{cb}^* V_{cd})$
(=Arg ($V_{ub}^*)$, in the standard  convention), the relative weak phase
between a CKM-favored $(b \to c)$ and a CKM-suppressed $(b \to u )$ decay
amplitude. This angle should be measured in a variety of ways
so as to check whether one consistently finds the same result.
There have been a lot of suggestions and discussions about
how to measure this quantity at $B$ factories \cite{ref3,ref4}. In Ref.
\cite{ref3} the authors proposed to extract $\gamma$ using the independent
measurements of $B \to D^0 K$, $B \to \bar D^0 K$ and $B \to D_{CP}^0 K$.
However, the charged $B$ meson decay mode $(B^- \to \bar D^0 K^-)$ is
difficult to measure experimentally. The reason is that the final
$\bar D^0$ meson should be identified using $\bar D^0 \to K^+ \pi^-$
but it is difficult to distinguish it from the doubly Cabibbo suppressed
$D^0 \to K^+ \pi^-$.
There are various methods to overcome these difficulties. In
Ref. \cite{ref5} Atwood et al
used different final states into which the
neutral $D$ meson decays, to extract information on $\gamma$. In Ref.
\cite{ref6} Gronau proposed that the angle $\gamma$ can be determined
by using the color allowed decay modes $B^- \to D^0 K^-$, $B^- \to
D_{CP}^0 K^- $ and their charge conjugation modes.
In Ref. \cite{ref7} a new method, using the isospin relations, 
is suggested to extract $\gamma$
by exploiting the decay modes $B \to D K^{(*)}$ that are not
Cabibbo suppressed.
Falk and Petrov \cite{ref9} recently proposed a new method for
measuring $\gamma$ using the partial rates for $CP$-tagged $B_s$
decays. It has been discussed in Ref. \cite{ref71} that it is possible
to extract $\gamma $ cleanly from $B_c \to D^0 D_s$ decays.
Sometime ago it was
pointed out in \cite{roy} that a clean extraction of the angle $\gamma $
is possible by studying the time dependence of the color allowed decays
$ B_s^0 (\bar B_s^0) \to D_s^\pm K^\mp $.

The angle $\gamma$
can also be measured using the $SU(3)$ relations between $B \to \pi K,
\pi\pi $ decay amplitudes \cite{ref10}. These analyses require additional
theoretical input, such as $SU(3)$ flavor symmetry and arguments for
the dynamical suppression of rescattering processes. While the validity
of some assumptions can be checked in the data itself, these approaches
leave theoretical uncertainties, which are hard to quantify reliably.
This limits the precision  with which they can be used to extract $\gamma $
by themselves. Nevertheless, they will provide important
cross-checks on other techniques as well as help us to address the discrete
ambiguities which theoretically cleaner methods leave unresolved.

It has been known for many years now that it is possible to cleanly
extract weak phase information using CP violating rate asymmetries in the
$B$ system. The earliest studies of such rate asymmetries concentrated on the
final states which are CP eigenstates. However, it soon became clear that
certain non-CP eigenstates can also be used. It has been shown by
Aleksan, Dunietz, Kayser and Le Diberder (ADKL)
\cite{roy1} that the CKM phase
can be cleanly determined in the $B$ decays to almost any final state
which is accessible to both $B_d^0 $ and $ \bar B_d^0$. In this paper we will
emphasize that the decay modes $B_s^0 \to f $ (where $f= \bar{D^0} \phi $)
and $\bar{ B_s^0} \to f$
can be used to probe the weak phase $\gamma $. In this case sizable
CP violating effect can occur because the two interfering amplitudes
$B_s^0 \to f$ and $\bar{B_s^0} \to f $ are of comparable size. 
Another nice feature of the above decay modes
is that unique weak phases are involved which arise from tree diagram
alone. No contamination from other weak phases are possible. Neither
penguin diagrams nor rescattering with different weak phases
exist. Hence the extracted weak phase is truely $\gamma $.
It has been discussed by Gronau and London \cite{gn91} that it is possible
to determine $\gamma $ considering
the time dependent decay rates of the three processes $B_s^0 \to D^0 \phi,
\bar{D^0} \phi $  and $D_1^0 \phi $ (where $D_1^0 $ is a neutral $D$ meson CP
eigenstate).

Here, we consider the final state $\bar D^0 \phi $ and the corresponding
vector vector modes $\bar D^{*0} \phi $ to which both $B_s^0$
and $\bar B_s^0 $ can decay. For the $\bar{D^0} \phi $ ($PV$)
final state we follow the ADKL method  \cite{roy1}
and for the vector-vector $(VV)$ final state we use the  
approach of Ref. \cite{nita00}. Since currently running
$e^+ e^- $ $B$ factories  operating at the $\Upsilon (4S)$ resonance will not
be in a position to explore $B_s$ decays, a strong emphasis has been given
to non-strange $B$ mesons in the recent literature.
However, the $B_s$ system also provides interesting strategies to determine
$\gamma $ and can prove to be useful system to understand CP violation.
So in order to make use of these methods and explore the CP violation in 
$B_s$-system, dedicated $B$ physics
experiments at hadron machines, such as LHC, BTeV etc.,
are the natural place. Within the
Standard Model, the weak $B_s^0 - \bar{B_s^0} $ mixing phase is very
small, and studies of $B_s$ decays involve very rapid $B_s^0 - \bar{B_s^0}$
oscillations due to large mass difference
$\Delta M_s = M_H^s - M_L^s $ between the mass eigenstates.
Future $B$-physics experiments performed at hadron machines should
be in a position to resolve the oscillations.

The paper is organised as follows. We present the method for the
determination of the angle $\gamma $ from the decay mode
$B_s^0(t)(\bar B_s^0(t)) \to \bar D^0 \phi $ in section II and from
$B_s^0(t)(\bar B_s^0(t)) \to \bar D^{*0} \phi $ in section III. Section IV
contains our conclusion.

\section{$\gamma $ from $B_s^0(\bar B_s^0) \to \bar D^0 \phi $}

Here we consider the final state $\bar{D^0} \phi $ to which both 
$B_s^0$ and $\bar{B_s^0}$ can decay. Both the amplitudes proceed via the
color suppressed tree diagrams only and there will be no penguin
contributions. The amplitude for $B_s^0 \to \bar D^0 \phi $
arises from the quark transition $\bar b \to \bar c u \bar s $
and has no weak phase in the Wolfenstein parametrization,
while the amplitude $\bar{B_s^0} \to \bar D^0 \phi $
arises from $ b \to u \bar c s $ and carries the weak phase
$e^{-i \gamma }$. The amplitudes also have the strong phases
$e^{i \delta_1}$ and $e^{i \delta_2}$. Thus, in general, one can
write the decay amplitudes as
\begin{eqnarray}
&& A(f) = {\rm Amp}~(B_s^0 \to \bar D^0 \phi)= A_1e^{i \delta_1}\nonumber\\
&& \bar A(f) = {\rm Amp}~(\bar B_s^0 \to \bar D^0 \phi)= A_2
e^{-i \gamma } e^{i \delta_2}\;.
\end{eqnarray}
The amplitudes for corresponding CP conjugate processes are given as
\begin{eqnarray}
&& \bar A(\bar f) = {\rm Amp}~(\bar B_s^0 \to D^0 \phi)=
A_1e^{i \delta_1}\nonumber\\
&& A(\bar f) = {\rm Amp}~(B_s^0 \to D^0 \phi)= A_2 e^{i \gamma}
e^{i \delta_2}\;.
\end{eqnarray}

Due to $B_s^0 -\bar B_s^0 $ mixing, a state which is created as a $B_s^0$
or a $\bar B_s^0$ will evolve in time into a mixture of both states
\cite{khoz99}. The weak phase of $B_s^0 -\bar B_s^0 $ mixing 
is decribed by the parameter $q/p$.
Within the Standard Model, $q/p = (V_{tb}^* V_{ts}/V_{tb} V_{ts}^*)$
is an excellent approximation. In the usual Wolfenstein parametrization
it has zero phase and $|q/p| \sim 1 $.
Since the final state $f = \bar D^0 \phi $ can be fed from
both $B_s^0 $ and $\bar B_s^0$,
the time dependent rates can be written as
\cite{khoz99}
\begin{eqnarray}
\Gamma (B_s^0(t) \to f) &=& \frac{ e^{-\Gamma t}}{2}~
 \biggr\{ \left [|A(f)|^2+|\bar A(f)|^2 \right ]
 \cosh \frac{\Delta \Gamma t}{2}
+\left [|A(f)|^2-|\bar A(f)|^2 \right ]\cos \Delta m t \nonumber\\
&+&2 {\rm Re}  \left [A(f)^* \bar A(f)\right ]
\sinh \frac{\Delta \Gamma t}{2}-2{\rm Im} \left [A(f)^* \bar A(f)
\right ]
\sin \Delta m t\biggr\}\;,
\nonumber\\
\Gamma (B_s^0(t) \to \bar f)&=& \frac{ e^{-\Gamma t}}{2}~
\biggr\{\left [|\bar A(\bar f)|^2+| A(\bar f)|^2\right ]
\cosh \frac{\Delta \Gamma t}{2}
-\left [|\bar A(\bar f)|^2-| A(\bar f)|^2\right ] \cos \Delta m t
 \nonumber\\
& +& 2 {\rm Re} \left [\bar A(\bar f)^*  A(\bar f)\right ]
\sinh \frac{\Delta \Gamma t}{2}+ 2 {\rm Im}
\left [\bar A(\bar f)^*  A(\bar f)\right ]
 \sin \Delta m t\biggr\}\;,
\nonumber\\
\Gamma (\bar B_s^0(t) \to \bar f)&= &\frac { e^{-\Gamma t}}{2}~
\biggr\{ \left [|\bar A(\bar f)|^2+|A(\bar f)|^2\right ]
 \cosh \frac{\Delta \Gamma t}{2}
+ \left [|\bar A(\bar f)|^2-|A(\bar f)|^2\right ]\cos \Delta m t
 \nonumber\\
& +& 2 {\rm Re} \left [\bar A(\bar f)^*  A(\bar f)\right ]
\sinh \frac{\Delta \Gamma t}{2}- 2 {\rm Im} \left [\bar A(\bar f)^*
A(\bar f)\right ]
 \sin \Delta m t\biggr\}\;,
\nonumber\\
 \Gamma (\bar B_s^0(t) \to f)&=& \frac{e^{-\Gamma t}}{2}~
 \biggr\{\left [|A(f)|^2+|\bar A(f)|^2\right ]
 \cosh \frac{\Delta \Gamma t}{2}
-\left [|A(f)|^2-|\bar A(f)|^2\right ] \cos \Delta m t \nonumber\\
&+&2 {\rm Re}\left [A(f)^* \bar A(f)\right ]
\sinh \frac{\Delta \Gamma t}{2}+2 {\rm Im}
\left [A(f)^* \bar A(f)\right ]
 \sin \Delta m t\biggr\}\;,
\end{eqnarray}
where $\Gamma $, $\Delta m$ and $\Delta \Gamma $ denote the average
width, the differences in mass and widths of the heavy and light
$B_s$ mesons respectively. If we denote the masses and widths of the two mass
eigenstates by $M_{L,H}$ and $\Gamma_{L, H}$ then we have
\begin{equation}
\Gamma = \frac{1}{\tau_{B_s}}=\frac{\Gamma_H+\Gamma_L}{2},
~~~~~~~~~\Delta m = M_H-M_L~~~~~~{\rm and}~~~~~~~~
\Delta \Gamma = \Gamma_H -\Gamma_L\;.
\end{equation}

Thus the time dependent measurement of $B_s^0(t)
\to f $ decay rates allow one to obtain
the following observables :
\begin{equation}
|A(f)|^2+|\bar A(f)|^2,~~~~ |A(f)|^2-|\bar A(f)|^2,~~~~
{\rm Re}[A(f)^* \bar A(f)]~~{\rm and}~~{\rm Im}[A(f)^* \bar A(f)]
\end{equation}
If $\Delta \Gamma =0$, then only the observables
$|A(f)|^2+|\bar A(f)|^2$, $ |A(f)|^2-|\bar A(f)|^2$ and
${\rm Im}(A(f)^* \bar A(f)) $ can be extracted from the time dependent study.
However, if $\Delta \Gamma $ is significantly different from zero
then all the four observables can be extracted.

Similarly, the time dependent measurements of $B_s^0(t) \to \bar f$
decay rates will give another four observables.
From these observables, the weak phase $\gamma $ can be determined, which
will be explained below. Now if
we substitute the decay amplitudes,
defined earlier in Eqs. (1) and (2), in Eq. (3) then
we get the decay rates as 
\begin{eqnarray}
\Gamma (B_s^0(t)  & \to & f) = \frac{ e^{-\Gamma t}}{2} ~
 \biggr\{ (|A_1|^2+|A_2|^2) \cosh \frac{\Delta \Gamma t}{2}
+(|A_1|^2-| A_2|^2 )\cos \Delta m t \nonumber\\
&+&2 A_1  A_2 \cos (\delta -\gamma)
\sinh \frac{\Delta \Gamma t}{2}-2  A_1  A_2 \sin (\delta -\gamma)
\sin \Delta m t\biggr\}\;,
\nonumber\\
\Gamma (B_s^0(t) & \to & \bar f)= \frac{ e^{-\Gamma t}}{2}~
\biggr\{(|A_1|^2+| A_2|^2)
\cosh \frac{\Delta \Gamma t}{2}
-(|A_1|^2-| A_2|^2) \cos \Delta m t
 \nonumber\\
& +& 2 A_1  A_2 \cos (\delta +\gamma)
\sinh \frac{\Delta \Gamma t}{2}+ 2  A_1  A_2 \sin (\delta +\gamma)
 \sin \Delta m t\biggr\}\;,
\nonumber\\
\Gamma (\bar B_s^0(t)  & \to & \bar f)= \frac { e^{-\Gamma t}}{2}~
\biggr\{ (|A_1|^2+|A_2|^2)
 \cosh \frac{\Delta \Gamma t}{2}
+ (|A_1|^2-|A_2|^2)\cos \Delta m t
\nonumber\\
&+& 2  A_1  A_2 \cos( \delta +\gamma)
\sinh \frac{\Delta \Gamma t}{2}-2 A_1  A_2 \sin (\delta +\gamma)
 \sin \Delta m t\biggr\}\;,
\nonumber\\
 \Gamma (\bar B_s^0(t)& \to & f)= \frac{e^{-\Gamma t}}{2}~
 \biggr\{(|A_1|^2+|A_2|^2)  \cosh \frac{\Delta \Gamma t}{2}
-(|A_1|^2-|A_2|^2) \cos \Delta m t \nonumber\\
&+&2  A_1  A_2 \cos (\delta -\gamma)
\sinh \frac{\Delta \Gamma t}{2}+2  A_1  A_2 \sin (\delta -\gamma)
 \sin \Delta m t\biggr\}\;,
\end{eqnarray}
where $\delta = \delta_2-\delta_1$ is the strong phase difference between
the two amplitudes $\bar{B_s^0}\to f$ and $B_s^0 \to f $.
Thus through the measurements of the time dependent rates,
it is possible to
measure the amplitudes $A_1$ and $A_2$ and the CP violating
quantities $S \equiv \sin(\delta+\gamma)$ and $\bar S \equiv \sin
(\delta - \gamma) $. In turn these
quantities will determine $\sin^2\gamma $ up to a four fold ambiguity
via the expression
\begin{equation}
\sin^2 \gamma = \frac{1}{2}\biggr[ 1-S \bar S \pm \sqrt{(1-S^2)
(1-\bar S^2}) \biggr]\;.\label{eq:t3}
\end{equation}
Here one of the signs on the right hand side gives the true $\sin^2
\gamma $, while the other gives $\cos^2 \delta $.
Thus $\sin^2  \gamma $ can be extracted cleanly with no
hadronic uncertainties but with four fold discrete ambiguity.
If the two mass eigenstates have widths which differ enough to result
in measurable effect, it becomes possible to experimentally resolve
some of the ambiguities in the determination of $\sin^2 \gamma $.

Alternatively, one can also determine $\sin 2 \gamma$ upto four
fold ambiguity  using the relation
\begin{equation}
\sin 2 \gamma =\biggr( S \sqrt{(1-\bar S^2)} - \bar S
\sqrt{(1-S^2)} \biggr)\;.\label{eq:t4}
\end{equation}
Thus the measurements allow one to determine both $\sin^2 \gamma $ and
$\sin 2 \gamma$ and hence to extract $\gamma $ unambiguously.

Now let us compute  the magnitudes of the amplitudes $A(f)$ and
$\bar A(f)$
and the corresponding branching ratios for such modes.
Using the factorization assumption the amplitudes for these modes are
given as
\begin{equation}
A(f) ={\rm Amp}( B_s^0(P) \to \bar D^0(q) \phi(\epsilon, p)) =
\frac{G_F}{\sqrt 2}  V_{cb}^* V_{us}~
a_2~ f_D~ 2 m_\phi~ A_0(m_D^2)~(\epsilon^* \cdot q)\;.\label{eq:r11}
\end{equation}
It should be noted here that the nonfactorizable
contributions play a significant role in colour suppressed
$B$ decays. However, since we are interested in
finding out the order of magnitudes of the decay amplitudes, here we
consider only the factorizable contributions. So in general,
one can expect that the estimated branching ratios may be either
enhanced or reduced due to the
nonfactorizable effects.

The decay amplitude
for the process $\bar{B_s^0} \to \bar{D^0} \phi $ can be
found by substituting the appropriate CKM matrix
elements in Eq. (\ref{eq:r11}).
Hence, one can write
\begin{equation}
|\bar A(f)/{A(f)}|=|{\rm Amp}(\bar B_s^0 \to \bar D^0 \phi)/{\rm Amp}
( B_s^0 \to \bar D^0
\phi)|=|{V_{ub}V_{cs}^*}/{V_{cb}^* V_{us}}| \sim 0.4\;. \label{eq:t2}
\end{equation}
Thus the advantage of using this final state is that since the two
interfering amplitudes are of comparable size the CP violating
asymmetry will be large.
The decay rates are given as
\begin{equation}
\Gamma( B_s^0 \to \bar D^0 \phi) = \frac{G_F^2}{4 \pi}  |V_{cb}^* V_{us}|^2
a_2^2 f_D^2 |A_0(m_D^2)|^2 |p|^3\;.
\end{equation}
The value of the form factor at zero momentum transfer, i.e., $A_0(0)$
can be found from BSW model \cite{bsw01} with value $A_0(0)=0.272$
\cite{xing01}. The momentum
dependence of $A_0$ can be found out assuming nearest pole dominance given as
\begin{equation}
A_0(q^2)=\frac{A_0(0)}{1-q^2/m_p^2}
\end{equation}
where the value of the pole mass $m_p$=5.38 GeV. Using $f_D=300 $ MeV
and $a_2=0.3$, the branching ratio is found to be
\begin{equation}
Br(B_s^0 \to \bar D^0 \phi)= 1.65 \times 10^{-5}\;.\label{eq:t1}
\end{equation}
Similarly using Eqs. (\ref{eq:t2}) and (\ref{eq:t1}),
the branching ratio for $\bar{B_s^0} \to \bar{D^0} \phi $
is found to be
\begin{equation}
Br(\bar{B_s^0} \to \bar D^0 \phi)= 2.64 \times 10^{-6}\;.
\end{equation}
The predicted branching ratios for these decay modes, which are
of the order of $\sim (10^{-5}-10^{-6})$
imply that they can easily be accessible in the second generation
hadronic $B$ factories. In hadron
machines such as Run II of Tevatron, LHC etc,
one anticipates a very large data sample of $B_s$ mesons.

We now make a crude estimate of the number of events required to measure
$\gamma $ by this method. An estimate of the sensitivity of the method can be
obtained by comparing it for example to $B_s^0 \to D_s^- K^+ $ decay.
The BTeV experiment with luminosity $2 \times 10^{32}~
{\rm cm}^{-2} {\rm s}^{-1}$ will produce $5 \times 10^{10}$ number of
$B_s^0 \bar{B_s^0}$ pairs per $10^7 $sec of running time.
The expected number of $B_s^0 \to D_s K $ events produced per year
is 13100 \cite{btev99}. The branching ratio for $B_s^0 \to D_s^- K^+ $
is $2 \times 10^{-4} $, while the branching ratio for
$B_s^0 \to \bar D^0 \phi $ is one order magnitude smaller.
If we take the branching ratios as :
$BR (B_s^0 \to \bar{D^0} \phi)\sim 10^{-5}$,
$BR(\bar{D^0} \to K^+ \pi^-$
and $K^+ \pi^- \pi^+ \pi^- $) to be $0.12$ and $BR (\phi \to K^+ K^-)=0.5$,
the reconstruction efficiency to be 0.05, and the trigger effeciency
level  as 0.9, we expect around 1350 number of reconstructed
$B_s^0 \to \bar{D^0} \phi $ events per year. 
An important issue of using this method is that the tagging of the initial 
$B_s$ meson is required. If we assume the tagging
efficiency to be 0.70 \cite{btev99} we expect approximately 945 tagged
events per year. For the LHC experiments also
several tagging strategies have already been studied
successfully \cite{ball00}. 
It is expected that a complete description of the tagging studies
will be available by the time the second
generation hadron machines are in operation. Thus the question
of whether this analysis is possible rests on whether the time resolution
is sufficient to separate the three different time-dependent terms.

\section{$\gamma $ from $B_s^0 (\bar B_s^0) \to \bar D^{* 0} \phi $}

Now we consider the final state $f=\bar D^{0*} \phi$,
consisting of two vector
mesons to which both $B_s^0 $ and $\bar B_s^0$ can decay. Because the final
state does not have a well defined angular momentum, the final
state $\bar{D^{*0}}\phi $ cannot be a CP eigenstate. By examining the decay
products of $\bar{D^{*0}} \phi $, one can measure the various helicity
components of the final state. Since each helicity state corresponds
to a well defined CP, an angular analysis of $B_s^0 \to \bar{D^{*0}} \phi$,
allows one to extract the CKM phase cleanly.
Here we will follow similar approach to that of London et al. \cite{nita00}
for the extraction of $\gamma $.
Due to the interference between
different helicity states, there are enough independent observables
for the decays of $B_s^0 $ and  $\bar{B_s^0}$ to the common final state
$\bar{D^{*0}} \phi$, from which the angle $\gamma $ can be extracted cleanly.
If both the final state mesons subsequently decay into two
$J^P =0^- $ mesons, i.e., $\bar{D^{*0}} \to K^+ \pi^- $ and
$\phi \to K^+ K^- $, the amplitude can be expressed in the
linear polarization
basis $(A_\parallel, A_\perp ~ {\rm and}~ A_0) $ \cite{dighe96} as
\begin{eqnarray}
{\cal A} &=& {\rm Amp}(B_s^0 \to f)= A_0 x_0+A_\parallel x_\parallel+
i A_\perp x_\perp \;, \nonumber\\
{\cal A^\prime } &=& {\rm Amp}(\bar B_s^0 \to f)= A_0^\prime x_0+
A_\parallel^\prime x_\parallel-
i A_\perp^\prime x_\perp \;,\label{eq:eqn67}
\end{eqnarray}
where $x_\lambda~(\lambda = 0, \parallel, \perp) $
are the coefficients of the helicity amplitudes in the
linear polarization basis, depend only on the angles describing
the kinematics \cite{dighe96}.

Using CPT invariance one can also write the amplitudes for the corresponding
CP conjugate processes as
\begin{eqnarray}
\bar{\cal A} &=& {\rm Amp}(\bar{B_s^0} \to \bar f)= \bar{A_0} x_0+
\bar{A_\parallel} x_\parallel-
i \bar{A_\perp} x_\perp \;, \nonumber\\
\bar{\cal {A^\prime} } &=& {\rm Amp}( B_s^0 \to \bar f)= \bar A_0^\prime x_0+
\bar A_\parallel^\prime x_\parallel+
i \bar A_\perp^\prime x_\perp \;.\label{eq:eqn68}
\end{eqnarray}
With the above Eqs. (\ref{eq:eqn67}) and (\ref{eq:eqn68})
the time dependent
decay rate $B_s^0(t) \to f $ can be written as
\begin{equation}
\Gamma(B_s^0(t) \to f) = e^{-\Gamma t} \sum_{\lambda \leq \sigma}
\biggr( X_{\lambda \sigma}+  Y_{\lambda \sigma}\cos \Delta m t
-Z_{\lambda \sigma} \sin \Delta m t \biggr) x_{\lambda} x_{\sigma}\;.
\end{equation}
Thus, by performing a time dependent study and angular analysis of the decay
$B_s^0(t) \to f $, one can measure the observables $X_{\lambda \sigma}$,
$Y_{\lambda \sigma}$ and $Z_{\lambda \sigma}$. In terms of the helicity
amplitudes these observables can be expressed as  follows
\begin{eqnarray}
&& X_{\lambda \lambda}=\frac{|A_\lambda|^2+|A_{\lambda }^\prime|^2}{2}\;,
~~~~~~~~~~~~~~~~~~~~~Y_{\lambda \lambda}=
\frac{|A_\lambda|^2-|A_{\lambda }^\prime|^2}{2}
\nonumber\\
&& X_{\perp i}=- {\rm Im} \left (A_\perp A_i^* -A_\perp^\prime
{A_i^\prime}^*\right )\;,
~~~~~~~~~~
X_{\parallel 0}={\rm Re} \left (A_\parallel A_0^* +A_\parallel^\prime
{A_0^\prime}^*\right )
\nonumber\\
&& Y_{\perp i}=- {\rm Im} \left (A_\perp A_i^* +
A_\perp^\prime {A_i^\prime}^* 
\right )\;,
~~~~~~~~~~~
Y_{\parallel 0}={\rm Re} \left (A_\parallel A_0^* -A_\parallel^\prime
{A_0^\prime}^*\right )
\nonumber\\
&& Z_{\perp i}=- {\rm Re} \left (A_\perp^* A_i^\prime +A_i^*
A_\perp^\prime
\right )\;,
~~~~~~~~~~~
Z_{\perp \perp}=-{\rm Im} \left (A_\perp^* A_\perp^\prime
\right )
\nonumber\\
&& Z_{\parallel 0}={\rm Im} \left (A_\parallel^* A_0^\prime +A_0^*
{A_\parallel^
\prime}^*\right )\;,
~~~~~~~~~~~~~~
Z_{ii}={\rm Im} \left (A_i^* A_i^\prime \right )\label{eq:r7}
\end{eqnarray}
where $i= \{0, \parallel\}$.
Similarly, the decay rate for $B_s^0 \to \bar f $ can be given as

\begin{equation}
\Gamma(B_s^0(t) \to f) = e^{-\Gamma t} \sum_{\lambda \leq \sigma}
\biggr( \bar X_{\lambda \sigma}+  \bar Y_{\lambda \sigma}\cos \Delta m t
-\bar Z_{\lambda \sigma} \sin \Delta m t \biggr) x_{\lambda} x_{\sigma}\;.
\end{equation}
The expressions for the observables
$ \bar X_{\lambda, \sigma}$, $\bar Y_{\lambda, \sigma}$ and $\bar Z_{\lambda,
\sigma} $ are similar to those given in Eq. (\ref{eq:r7}) with the
replacements $ A_\lambda \to \bar A_\lambda^\prime $ and
$ A_\lambda^\prime \to \bar A_\lambda$.

For the purpose of the exctraction of $\gamma$ let us first write the helicity
amplitudes for each processes. Since they are described by the colour
suppressed tree diagrams, only a single CKM weak phase will be involved
in these amplitudes and are given as

\begin{eqnarray}
&& A_\lambda = {\rm Amp} (B_s^0 \to f)_\lambda= a_\lambda ~
e^{i \delta_\lambda^1}\nonumber\\
&& A_\lambda^\prime ={\rm Amp} (\bar B_s^0 \to f)_\lambda= b_\lambda ~
e^{-i \gamma } ~ e^{i \delta_\lambda^2}\nonumber\\
&& \bar A_\lambda^\prime = {\rm Amp} ( B_s^0 \to \bar f)_\lambda=
b_\lambda ~
e^{i \gamma }~ e^{i \delta_\lambda^2}\nonumber\\
&& \bar A_\lambda = {\rm Amp} (\bar B_s^0 \to \bar f)_\lambda=
a_\lambda ~ e^{i \delta_\lambda^1}\;,
\end{eqnarray}
where $\gamma $ = Arg$(V_{ub}^*)$ represents the weak phase and
$\delta_\lambda^1$, $\delta_\lambda^2$ are the strong phases.
With these above expressions for the various amplitudes, we now show how to
extract the weak phase $\gamma$ using the decay rate measurements. It
is now very easy to see that the observables can be written in terms of the
helicity amplitudes as
\begin{equation}
X_{\lambda \lambda}=\bar X_{\lambda \lambda}=\frac{|a_\lambda|^2
+|b_\lambda|^2}{2}\;,
~~~~~~~~~~~~~Y_{\lambda \lambda}=-\bar Y_{\lambda \lambda}
=\frac{|a_\lambda|^2-|b_{\lambda }|^2}{2} \label{eq:a7}\;.
\end{equation}
Thus, one can determine the magnitudes of various helicity amplitudes
$|a_\lambda|^2$ and $|b_\lambda|^2$ from Eq. (\ref{eq:a7}).
Next we obtain the expressions for the observables : $X_{\perp i}$,
$Y_{\perp i}$, $X_{\parallel 0}$ and $Y_{\parallel 0}$,
\begin{eqnarray}
X_{\perp i} = -\bar X_{\perp i}=b_{\perp}b_i\sin (\delta_{\perp}
+\Delta_i-\delta_i)-a_{\perp }a_i \sin \Delta_i \nonumber\\
Y_{\perp i} = \bar Y_{\perp i}=-b_{\perp}b_i\sin (\delta_{\perp}
+\Delta_i-\delta_i)-a_{\perp }a_i \sin \Delta_i\label{eq:r1}
\end{eqnarray}
where $\Delta_i=\delta_\perp^1-\delta_i^1 $ and
$\delta_\lambda=\delta_\lambda^2-\delta_\lambda^1 $. 
Using Eq. (\ref{eq:r1}) one can solve for $a_{\perp }a_i \sin \Delta_i$.
Similarly, one can write
\begin{eqnarray}
&&X_{\parallel 0} = \bar X_{\parallel 0}=b_{\parallel}b_0\cos
(\delta_{\parallel}
+\Delta-\delta_0)+a_{\parallel}a_0 \cos \Delta \nonumber\\
&&Y_{\parallel 0} =- \bar Y_{\parallel 0}=-b_{\parallel}b_0
\cos (\delta_{\parallel}
+\Delta-\delta_0)+a_{\parallel}a_0 \cos \Delta \;,\label{eq:r2}
\end{eqnarray}
where $\Delta=\delta_\parallel^1-\delta_0^1 $. Thus
one can solve for $a_{\parallel }a_0 \cos
\Delta$ using Eq. (\ref{eq:r2})

The coefficients of $\sin (\Delta m t )$ term, which can be obtained
in a time dependent study, can be written as
\begin{eqnarray}
&&Z_{ii}=a_i b_i \sin(\delta_i-\gamma)\;,
~~~~~~~~~~~~~\bar{Z_{ii}}=-a_i b_i \sin(\delta_i+\gamma)\;,\nonumber\\
&&Z_{\perp \perp}= -a_\perp b_\perp \sin(\delta_\perp-\gamma)\;,
~~~~~~~\bar{Z}_{\perp \perp}= a_\perp b_\perp \sin(\delta_\perp+\gamma)
\;.
\end{eqnarray}
Thus we can find
\begin{eqnarray}
&& 2b_i \cos \delta_i=- \frac{Z_{ii}+\bar Z_{ii}}{a_i \sin \gamma}\;,
~~~~~~~~~
2b_i \sin \delta_i= \frac{Z_{ii}-\bar Z_{ii}}{a_i \cos \gamma}\;,
\nonumber\\
&& 2b_\perp \cos \delta_\perp= \frac{Z_{\perp \perp}+\bar Z_{
\perp\perp}}{a_\perp \sin \gamma}\;,~~~~~~~~
2b_\perp \sin \delta_\perp= -\frac{Z_{\perp\perp}-\bar Z_{
\perp\perp}}{a_\perp \cos \gamma}\;.\label{eq:a8}
\end{eqnarray}
Similarly, the terms involving interference of different helicities are
given as
\begin{eqnarray}
Z_{\perp i} =-\biggr[a_\perp b_i \cos (\delta_i -\Delta_i-\gamma)
+a_i b_\perp \cos(\delta_\perp + \Delta _i -\gamma)\biggr]\nonumber\\
\bar Z_{\perp i} =-\biggr[ a_\perp b_i  \cos (\delta_i -\Delta_i+\gamma)
+a_i b_\perp \cos(\delta_\perp + \Delta _i +\gamma)\biggr]\label{eq:a9}
\end{eqnarray}
and
\begin{eqnarray}
&&Z_{\parallel 0} =\biggr[a_\parallel b_0 \sin (\delta_0 -\Delta-\gamma)
+a_0 b_\parallel \sin(\delta_\parallel + \Delta  -\gamma)\biggr]\nonumber\\
&&\bar Z_{\parallel 0} =-\biggr[ a_\parallel b_0  \sin
(\delta_0 -\Delta+\gamma)
+a_0 b_\parallel \sin(\delta_\parallel + \Delta
 +\gamma)\biggr]\label{eq:a10}
\end{eqnarray}
Considering all the above information together, we are now in a position
to extract the weak phase $\gamma $. Making use of Eq. (\ref{eq:a8}) we can
rewrite Eq. (\ref{eq:a9}) in the following two useful forms:

\begin{eqnarray}
 Z_{\perp i}+\bar Z_{\perp i}&=&a_i a_\perp \cos \Delta_i
 \cot \gamma
 \biggr[ \frac{Z_{ii}+\bar Z_{ii}}{a_i^2}  -\frac{Z_{\perp \perp}
 +\bar Z_{\perp \perp}}{a_\perp^2}\biggr]\nonumber\\
&-& a_i a_\perp \sin \Delta_i
 \biggr[ \frac{Z_{ii}-\bar Z_{ii}}{a_i^2}  +\frac{Z_{\perp \perp}
 -\bar Z_{\perp \perp}}{a_\perp^2}\biggr]\label{eq:a11}
 \end{eqnarray}
 \begin{eqnarray}
Z_{\perp i}-\bar Z_{\perp i}&=&- a_i a_\perp \cos \Delta_i \tan \gamma
 \biggr[ \frac{Z_{ii}-\bar Z_{ii}}{a_i^2}  -\frac{Z_{\perp \perp}
-\bar Z_{\perp \perp}}{a_\perp^2}\biggr]\nonumber\\
&-& a_i a_\perp \sin \Delta_i
 \biggr[ \frac{Z_{ii}+\bar Z_{ii}}{a_i^2}  +\frac{Z_{\perp \perp}
+\bar Z_{\perp \perp}}{a_\perp^2}\biggr] \label{eq:a12}
\end{eqnarray}
Now let us closely look into the terms involved in the above two
Eqs. (\ref{eq:a11}) and (\ref{eq:a12}).
We already know most of the terms : {\it i.)} $Z_{\lambda \lambda}$,
$\bar{Z}_{\lambda \lambda}$ are measurable quantities.
{\it ii.)} $a_\lambda^2$ are known quantities, can be determined from Eq.
(\ref{eq:a7}). {\it iii.)} $a_i a_\perp \sin \Delta_i $ is obtained
from Eq. (\ref{eq:r1}). Thus, these two Eqs. involve only
two unknown quantities $\tan \gamma $ and $a_i a_\perp \cos \Delta_i $
and hence can be easily solved upto a sign ambiguity in each of these
quantities. Thus $\tan^2 \gamma $ or equivalently $\sin^2 \gamma $
can be obtained from the angular analysis.

Similarly, if we put all the informations in Eq. (\ref{eq:a10}) we obtain
the following relations
\begin{eqnarray}
 Z_{\parallel 0}+\bar Z_{\parallel 0}&=& a_0
 a_{\parallel} \cos \Delta \cot \gamma
 \biggr[ \frac{Z_{00}+\bar Z_{00}}{a_0^2}  +\frac{Z_
 {\parallel ~ \parallel}
 +\bar Z_{\parallel ~\parallel }}{a_{\parallel }^2}\biggr]\nonumber\\
&-& a_{\parallel } a_0 \sin \Delta
 \biggr[ \frac{Z_{00}-\bar Z_{00}}{a_0^2}-
 \frac{Z_{\parallel ~\parallel }- \bar Z_{\parallel~ \parallel }}{a_
 {\parallel }^2}\label{eq:a13}
\end{eqnarray}
\begin{eqnarray}
 Z_{\parallel 0}-\bar Z_{\parallel 0}&=& a_0
 a_{\parallel} \cos \Delta
 \biggr[ \frac{Z_{00}-\bar Z_{00}}{a_0^2}  +\frac{Z_
 {\parallel~ \parallel}
-\bar Z_{\parallel~ \parallel }}{a_{\parallel }^2}\biggr]\nonumber\\
&+& a_{\parallel } a_0 \sin \Delta \cot \gamma
 \biggr[ \frac{Z_{00}+\bar Z_{00}}{a_0^2}-
 \frac{Z_{\parallel~ \parallel }+ \bar Z_{\parallel~ \parallel }}{a_
 {\parallel }^2}\label{eq:a14}
\end{eqnarray}
In the above two Eqs. (\ref{eq:a13}) and (\ref{eq:a14}) we know all the
quantities except two i.e. $a_\parallel  a_0 \cos \Delta $ and $\cot
\gamma $. Thus these two unknowns can easily be determined from these
two equations. Thus one can solve for $\tan^2 \gamma $ or equivalently
$\sin^2 \gamma$ without any hadronic uncertainties.

To calculate the number of events necessary to determine $\gamma $ using
this method we have to know the branching ratios for the decay modes
$B_s^0 (\bar{B_s^0}) \to \bar{D^{*0} }\phi $. The decay widths
(in units of $10^{12} s^{-1} $)
for these decay modes are calculated in Ref. \cite{gatto93} as
\begin{equation}
\Gamma(B_s^0 \to \bar{D^{*0} }\phi )= 2\times a_2^2 |V_{cb}^* V_{us}|^2\;,
~~~~~~~~~~~~~
\Gamma(\bar{B_s^0} \to \bar{D^{*0}} \phi )= 1.9 a_2^2 |V_{ub} V_{cs}^*|^2\;.
\end{equation}
Thus the branching ratios are given as
\begin{equation}
BR(B_s^0 \to \bar{D^{*0} }\phi )= 2.09 \times 10^{-5}\;,
~~~~~~~~~~~~~
BR(\bar{B_s^0} \to \bar{D^{*0}} \phi )= 3.61\times 10^{-6}\;.
\end{equation}
Thus if we assume the same selection method as we have done for $B_s^0
\to \bar{D^0} \phi $ case, here we get approximately 2700
reconstructed $B_s^0 \to \bar{D^{*0} }\phi $ events per year of running
at BTeV. Assuming the tagging efficiency to be 0.70, one expects
approximately 1890 tagged events per year.

\section{Conclusion}

In this paper we have discussed the determination of the CKM phase $\gamma $
from time dependent measurement of the pure tree nonleptonic
$B_s^0 $ decay modes
$B_s^0(t)(\bar B_s^0(t)) \to \bar D^0 \phi $ and
$B_s^0(t)(\bar B_s^0(t)) \to \bar D^{*0} \phi $.
For the former case we have
used the formalism similar to that of ADKL method \cite{roy1}
and for the latter case we have followed
the approach of London et al \cite{nita00}.
The advantage of these decay modes is that they are described by pure tree diagrams
and hence free from theoretical hadronic uncertainties. 
Within the Standard Model these modes are expected to exhibit
branching ratios at the $10^{-5}-10^{-6}$ level. So they
are expected to be easily accessible in
the second generation hadronic $B$ factories.

The accurate determination of phase $\gamma$ of the 
unitarity triangle is really challenging.
Therefore, one should study as many decay modes as possible to
cross check other findings and also explore various
strategies for the clean determination of it.

For the case of ($PV$) final states i.e,
$B_s^0(\bar B_s^0) \to \bar D^0 \phi$ we have obtained four observables
($A_1$, $A_2$, $S$ and $\bar S$), from the corresponding time dependent
decay rates. From these observables $\gamma $ can be extracted using
Eq. (\ref{eq:t3}) or (\ref{eq:t4}).
Furthermore, one can also find the information about
strong phase (Eq. (\ref{eq:t3})) from this analysis.
Next, we have considered the (VV)  final states
$B_s^0(\bar B_s^0) \to \bar D^{*0} \phi$.
We have used linear polarization basis to write the
decay rates in terms of the observables ($X$, $Y$, $Z$).
From these observables, one can exctract $\gamma$ solving
Eqs. (28, 29) or Eqs. (30, 31).
It should be emphasized here that using our analysis,
the exctraction of $\gamma$ can be done cleanly
without any hadronic uncertainties in both the cases
of $PV$ and $VV$ final states of $B_s$ meson,
but with some amout of discrete ambiguities. Furthermore, these
modes may possibly guide us to know physics beyond Standard Model and/or
valuable informations regarding the nature of CP violation.

To summarize, we point out here that it is indeed possible to determine the weak phase $\gamma$ 
cleanly from the  time dependent measurement of the nonleptonic decay modes
$B_s^0(\bar B_s^0) \to \bar D^0 \phi $ and the corresponding vector vector
modes $ B_s^0(\bar B_s^0) \to \bar D^{* 0} \phi $. 
The strategies presented in this paper appear
to be particularly interesting for second generation experiments
at hadron  machines such as LHC and BTeV, where also the very
powerful physics potential of the $B_s$ meson can be exploited.

\section{Acknowledgements}
We are grateful Professor S. Stone for useful comments and for
providing us valuable information
concerning the experimental feasibility of such modes at BTeV
experiment. AKG and MPK would like to thank Council of Scientific and
Industrial Research, Government of India, for financial support.

\end{document}